\documentclass[a4paper,leqno]{llncs}
\usepackage[british]{babel}

\usepackage{amsmath}
\usepackage{amssymb}

\usepackage{enumerate}
\usepackage{xspace}
\newcommand{\mathcmd}[1]{\ensuremath{#1}\xspace}
\newcommand{\mlfont}{\textsl}
\newcommand{\ml}[1]{\mathcmd{\mlfont{\upshape #1}}}
\newcommand{\dlfont}{\mathcal}
\newcommand{\dl}[1]{\mathcmd{\dlfont{#1}}}
\newcommand{\K}{\ml{K}}
\newcommand{\ALC}{\dl{ALC}}
\newcommand{\idRole}{\mathcmd{\textsf{\upshape\mdseries id}}}
\newcommand{\ALBO}{\dl{ALBO}}
\newcommand{\ALBOid}{\dl{ALBO\smash[t]{\mbox{${}^{\idRole}$}}}}
\def\lAnd{\wedge}
\def\lOr{\vee}
\def\lImp{\rightarrow}
\def\lNot{\neg}
\def\lEq{\leftrightarrow}

\let\Or\lOr

\let\Not\lNot

\let\pBox\lBox
\let\at\lAt
\let\bigAnd\lBigAnd
\let\bigOr\lBigOr
\makeatletter
\def\@@disp#1#2#3#4{\@ifnextchar[{\@@ydisp{#1}{#2}{#3}{#4}}{\@@xdisp{#1}{#2}{#3}{#4}}}
\def\@@xdisp#1#2#3#4{\@spbegintheorem{#2}{#1}{#3}{#4}\ignorespaces}
\def\@@ydisp#1#2#3#4[#5]{\@spopargbegintheorem{#2}{#1}{#5}{#3}{#4}\ignorespaces}
\newcommand{\spnewrefthm}[4]{%
\newenvironment{#1}[1]{%
   \@@disp{\ref{##1}}{#2}{#3}{#4}%
}{\@endtheorem}
}%
\makeatother
\spnewrefthm{reftheorem}{Theorem}{\bfseries}{\itshape}
\newcommand{\metasmbfont}{\mathsf}

\newcommand{\seq}[1]{\overline{#1}}
\newcommand{\ecl}[1]{\|#1\|}
\let\Lang=\lang
\makeatletter
\def\@define#1{%
    \mathcmd{%
        \stackrel%
            {\mbox{%
                \tiny\ensuremath{\metasmbfont{def}}%
                  }%
            }%
            {{#1}}%
            }%
              }
\def\sm@shdefine#1{%
\smash[t]{\@define{#1}}%
}
\newcommand{\define}{%
        \mathchoice{\@define{\ =\ }}{\sm@shdefine{=}}{\sm@shdefine{=}}{\sm@shdefine{=}}%
}
\newcommand{\defiff}{%
        \mathchoice%
            {\@define{\ \Longleftrightarrow\ }}%
            {\sm@shdefine{\Longleftrightarrow}}%
            {\sm@shdefine{\Leftrightarrow}}%
            {\sm@shdefine{\Leftrightarrow}}%
}
\newcommand{\branch}[1]{\seg{#1}}
\newcommand{\seg}[1]{\mathcmd{\mathcal{#1}}}
\newcommand{\tor}{\,\mid\,}
\newcommand{\tand}{\,\,\ \ \,}

\let\tableaureffmt\@relax
\newcommand{\tableaulblfont}{\sffamily}
\iftagsleft@
\def\tableauleftlbldelim{}
\def\tableaurightlbldelim{:\ }
\else
\def\tableauleftlbldelim{\ :}
\def\tableaurightlbldelim{}
\fi
\def\tableaulblfmt#1{\text{\tableaulblfont\tableauleftlbldelim #1\tableaurightlbldelim}}
\def\@xtableaurule#1#2{%
\genfrac{}{}{}{0}{#1}{#2}%
}
\def\@ytableaurule#1#2[#3]{%
\def\lbl@{\tableaureffmt{#3}}
\iftagsleft@%
\tableaulblfmt{#3}\@xtableaurule{#1}{#2}%
\else%
\@xtableaurule{#1}{#2}\tableaulblfmt{#3}%
\fi%
\let\label\ltx@label
\let\@currentlabel\lbl@
}
\def\tableaurule#1#2{%
\@ifnextchar[{\@ytableaurule{#1}{#2}}{\@xtableaurule{#1}{#2}}%
}

\def\@xinlinetableaurule#1#2{%
{#1}/{#2}%
}
\def\@yinlinetableaurule#1#2[#3]{%
\def\lbl@{\tableaureffmt{#3}}
\iftagsleft@%
\tableaulblfmt{#3}\@xinlinetableaurule{#1}{#2}%
\else%
\@xinlinetableaurule{#1}{#2}\tableaulblfmt{#3}%
\fi%
\let\label\ltx@label
\let\@currentlabel\lbl@
}
\def\inlinetableaurule#1#2{%
\@ifnextchar[{\@yinlinetableaurule{#1}{#2}}{\@xinlinetableaurule{#1}{#2}}%
}

\newcount\@tskipcnt
\def\tfillsymbol{\mbox{\fontsize{3}{4}\selectfont.}}
\def\tfill{%
  \leavevmode
  \cleaders \hb@xt@ .44em{\hss{\tfillsymbol}\hss}\hfill
  \kern\z@}

\newcommand{\tbranch}{{\blacktriangleright}}

\def\@@tskip{\phantom{\tbranch}}%
\def\@@@tskip{\@@tskip\advance\@tskipcnt\m@ne}%
\def\@tskip[#1]{%
\@tskipcnt#1
\loop\ifnum\@tskipcnt>\z@ \@@@tskip\repeat}%

\def\tskip{%
\@ifnextchar[{\@tskip}{\@@tskip}}

\let\tableaureffmt\tableaulblfmt
\makeatother
\renewcommand{\tableauleftlbldelim}{(}
\renewcommand{\tableaurightlbldelim}{)~}
\renewcommand{\tableaureffmt}[1]{%
\text{\tableaulblfont\upshape({#1})}%
}
\newcommand{\inlineproverfont}{\scshape\mdseries}
\DeclareTextFontCommand{\proverfont}{\inlineproverfont}
\makeatletter
\newcommand{\@mettel}{{\inlineproverfont Met\kern-.063emTeL}}
\DeclareRobustCommand{\mettel}{\mathcmd{\text{\@mettel}}}
\def\@mettelII{{\inlineproverfont\@mettel{}\kern-.33em\raise.7ex\hbox{2}}}
\def\f@smash{\ht\z@\z@ \box\z@}
\newcommand{\@@mettelII}{\makesm@sh{\@mettelII}\f@smash}
\DeclareRobustCommand{\mettelII}{\mathcmd{\text{\@@mettelII}}}
\makeatother

\def\complexityfont{\sffamily\upshape}
\def\complexity#1{\mathcmd{\text{\complexityfont #1}}}
\def\ExpTime{\complexity{ExpTime}}

\newcommand{\inverse}{\mathord{\sim}}
\newcommand{\FO}{\textsf{\upshape FO}\xspace\/}

\newcommand{\Sorts}{\textsf{\upshape Sorts}\xspace\/}

\newcommand{\I}{\mathcal{I}}

\newcommand{\IB}{{\I(\branch{B})}}
\newcommand{\D}{\mathsf{D}}
\newcommand{\s}{\mathsf{s}}
\newcommand{\simB}{\mathop{\sim_\branch{B}}}
\newcommand{\formulae}{\mathsf{f}}
\newcommand{\relations}{\mathsf{r}}
\newcommand{\ar}{\mathsf{ar}}
\newcommand{\Rf}{\mathcmd{\mathop{\textsf{\upshape ref}}}}
\newcommand{\Rfr}{\Rf}

\newcommand{\Km}{{\mathcmd{\K_m}}}
\newcommand{\KmNot}{{\mathcmd{\Km(\Not)}}}

\begin{document}

\mainmatter  %

\title{Refinement in the Tableau Synthesis Framework}

\author{Dmitry Tishkovsky\and Renate A. Schmidt%
\thanks{This research is supported by UK EPSRC research grant EP/H043748/1.}}
\authorrunning{D. Tishkovsky, R. A. Schmidt}
\institute{School of Computer Science, The University of Manchester, UK
}
\toctitle{Refinement in the Tableau Synthesis Framework}
\tocauthor{D. Tishkovsky, R. A. Schmidt}
\maketitle

\begin{abstract}
This paper is concerned with the possibilities of refining and
improving calculi generated in the tableau synthesis framework~\cite{SchmidtTishkovsky-ASTC-lmcs-2011}. 
A general method in the tableau synthesis framework allows
to reduce the branching factor of tableau rules and preserves completeness
if a general rule refinement condition holds.
In this paper we consider two approaches to satisfy this general rule
refinement condition.
\end{abstract}

\section{Introduction}
The tableau method is one of the most popular deduction approaches in
automated reasoning. 
Tableau methods in various forms 
are successfully applied for many logics 
and are especially apt for dynamically developing areas
requiring new logical formalisms. 
However, developing a tableau calculus for a new logic 
is still a challenging task 
which usually involves tedious proving
of soundness and completeness results.

Based on the collective experience in the area,
in recent work~\cite{SchmidtTishkovsky-ASTC-lmcs-2011} we introduced
a general framework for synthesising and studying semantic tableau
calculi for propositional logics.
The framework formalises a three step process for transforming the
definition of the semantics of a logic into a sound and complete
tableau calculus.
The first two steps are to specify the semantics of the logic
and to extract tableau inference rules from the semantic
specification. 
Tableau rule extraction is 
automatic and produces a set of tableau rules operating
on formulae in a generic tableau language.
When certain natural conditions hold, the generated rules form a sound
and constructively complete tableau calculus.

Initially, the generated calculi are in a basic form.
Two deficiencies can be identified.
One is that some rules of the generated calculus 
can have branches which are not necessary for guaranteeing completeness.
The other problem is that the tableau language of the tableau synthesis framework
can be excessively laden with extra-logical notation. 
Often, but not always, the generated tableau calculus can be encoded more compactly
within the language of the logic.
Both problems decrease the performance of tableau algorithms
based on the calculus.
That is why the tableau synthesis framework, addressing these problems,
defines a third \emph{crucial} step: the refinement of the generated tableau calculus. 

In~\cite{SchmidtTishkovsky-ASTC-lmcs-2011}
we describe two refinements:
rule refinement, and a refinement that 
internalises the language of the calculus within the language of the logic.
While the internalisation of the tableau language
can be done routinely by extending and massaging the language of the logic
and retains constructive completeness of the calculus,
rule refinement, and whether rule refinement preserves
constructive completeness is generally more difficult to establish.
Currently, rule refinement requires 
the verification of a general rule refinement condition 
which is inductive and needs to be checked manually.
For the purposes of automating rule refinement
it is therefore important to find
other less generic conditions
sufficient to preserve constructive completeness
and, yet, can be automatically verified.

In this paper we describe two new approaches 
to satisfy the general rule
refinement condition and illustrate these approaches on the examples
of multi-modal logic $\Km$ satisfying any first-order frame conditions, and the
logic $\Km(\Not)$, which is an extension of $\Km$ with the negation
operator on accessibility relations. 
In the first approach we show
how to extend a set of non-refinable rules
by altering the semantic specification of the logic
and obtain a modified set of rules which can be refined.
For the second approach we present a special atomic rule refinement
condition and prove that it implies the general rule
refinement condition.
Consequently, this guarantees that the atomic rule refinement preserves constructive completeness
of the tableau calculus.

We identify two important cases when the atomic rule refinement
condition is satisfied automatically.
The first case allows to refine rules which are generated from frame
conditions for arbitrary combinations of modal logics.
In the second case, using the atomic rule refinement, we show how
to transform the generated tableau calculus into a hypertableau-like
calculus and prove that the transformation preserves 
constructive completeness of the calculus.

The paper is structured as follows.
The logics $\Km$ and $\KmNot$ which serve as running examples are introduced in Section~\ref{section:logics}.
We recall some notions from the tableau synthesis framework~\cite{SchmidtTishkovsky-ASTC-lmcs-2011}
and generate tableau calculi for the considered logics
in Section~\ref{section:framework}.
Refinements introduced in the tableau synthesis framework
are reviewed in Section~\ref{section:existing refinements}.
Examples of the application of rule refinements are given in Section~\ref{section:examples of refinement}.
Atomic rule refinement, a special case of the rule refinement,
is introduced and investigated in Section~\ref{section:atomic rule refinement}.
In Section~\ref{section:refinement for frame conditions}, 
we apply atomic rule refinement to 
the rules generated from frame conditions for the logic $\Km$.
We use the atomic rule refinement to construct
a hypertableau-like calculus for the logic $\KmNot$ in Section~\ref{section:hypertableau}.
We conclude with a discussion of the presented results in Section~\ref{section:discussion}.
For the benefit of reviewers proofs and technical details are included in the Appendix.

\section{The logics $\Km$ and $\KmNot$}\label{section:logics}
As examples to illustrate results of this paper
we consider 
two logics:
multi-modal logic $\Km$ with possible frame conditions,
and the logic $\KmNot$, the extension of $\Km$ 
with the operator of negation of relations.

Following the tableau synthesis framework~\cite{SchmidtTishkovsky-ASTC-lmcs-2011}
the languages $\Lang{L}(\Km)$ and $\Lang{L}(\KmNot)$ of these logics
have two sorts: a sort for formulae $\formulae$ and a sort for relations $\relations$.
The sorts of relations in these languages are formed over a 
set of relational constants $\{a_1,\ldots,a_m\}$.
In $\Lang{L}(\Km)$ every relation is a relational constant, and
in $\Lang{L}(\KmNot)$
every \emph{relation} $\alpha$ is defined by the BNF $\alpha\define a_1\mid \cdots\mid a_m\mid \Not\alpha$,
where $\Not$ is the operator of negation on relations. 
The sorts of formulae in both languages are formed over a set of propositional variables $\{p,q,\ldots\}$ and
every \emph{formula} $\phi$ of each language is defined 
by the BNF
$\phi\define p\mid \Not\phi\mid \phi\Or\phi\mid \pBox{\alpha}\phi$,
where $\alpha$ ranges over all relations of the language.

According to the tableau synthesis framework,
the semantic specification language $\FO(\Km)$ for $\Km$
is a (multi-sorted) first-order language over sorts of
$\Lang{L}(\Km)$ and an additional \emph{domain sort} $\D$.
Expressions of $\Lang{L}(\Km)$ are naturally embedded
into $\FO(\Km)$ as terms of appropriate sorts.
That is, every logical connective of $\Lang{L}(\Km)$
is a functional symbol of $\FO(\Km)$.
Every propositional variable of $\Lang{L}(\Km)$ is
an individual variable of the sort $\formulae$ in $\FO(\Km)$.
Besides the individual constants $a_1,\ldots,a_m$ for relations,
the language $\FO(\Km)$ has a countable set of relation variables $r,r',\ldots$.
The additional sort $\D$ %
has a countable set of individual variables $x,y,z,\ldots$. 
Furthermore, the semantic specification language has 
two predicate symbols $\nu_\formulae$ and $\nu_\relations$
of sorts $(\formulae,\D)$ and $(\relations,\D,\D)$, respectively.
The symbols $\nu_\formulae$ and $\nu_\relations$
are required for the purpose of representing
satisfiability with respect to domain elements.
The meaning of these symbols can be understood 
from definitions below.

The semantic specification of $\Km$ consists of the following three formulae. %
\begin{align*}
\forall x\; (\nu_\formulae(\Not p, x)&\lEq \lNot\nu_\formulae(p,x))\\
\forall x\; (\nu_\formulae(p\Or q), x)&\lEq \nu_\formulae(p,x)\lOr\nu_\formulae(q,x))\\
\forall x\; (\nu_\formulae(\pBox{r}p, x)&\lEq \forall y\;(\nu_\relations(r,x,y)\lImp\nu_\formulae(p,y)))
\end{align*}

A \emph{model} $\I$ of $\Km$ is a tuple
$\I=(\Delta^\I,\nu_\formulae^\I,\nu_\relations^\I)$
where $\Delta^\I$ is a non-empty set for interpretation
of variables of the domain sort,
$\nu_\formulae^\I$ and $\nu_\relations^\I$
are interpretations of the predicates 
$\nu_\formulae$ and $\nu_\relations$ respectively,
and all the formulae of the semantic specification for $\Km$ are true in $\I$
(for all $\Km$-formulae $p$, $q$, and $\Km$-relation $r$).
The purpose of symbols $\nu_\formulae$ and $\nu_\relations$ is to define the
semantics of the connectives of the logic by using conditions similar to satisfaction
conditions in standard definitions.
That is, given a $\Km$-model $\I$ %
and elements $v$ and $w$ from its domain, 
for a formula $\phi$
and a relation $\alpha$ of $\Lang{L}(\Km)$,
$\I\models\nu_\formulae(\phi,v)$ can be read as `$\phi$ is true in $v$' 
and  $\I\models\nu_\relations(\alpha,v,w)$ is understood as
`$w$ is an $\alpha$-successor of $v$ in $\I$'.

For illustrative purposes, in Section~\ref{section:refinement for frame conditions} we
consider two properties of the relations of $\Km$.
Following the terminology in modal logic we refer to such properties as \emph{frame conditions}.
The first condition expresses irreflexivity of relations
and is specified by the formula
$\forall x\,\lNot\nu_\relations(r,x,x)$.
The second frame condition is intentionally even more exotic.
It states existence of an immediate predecessor
for each world in the model and
is specified by the formula
\[\forall x\exists y\forall z\Bigl(\nu_\relations(r,y,x)\lAnd x\not\approx y\lAnd%
\bigr((\nu_\relations(r,y,z)\lAnd\nu_\relations(r,z,x))\lImp(z\approx x\lOr z\approx y)\bigl)\Bigr).\]

The semantic specification language $\FO(\KmNot)$ of the logic $\KmNot$
differs from the language $\FO(\Km)$ only in that the sort of relations
involves the negation operator.
The semantic specification of $\KmNot$ extends the semantic specification of $\Km$ by
the following $\FO(\KmNot)$-formula.
\[\forall x\forall y\; (\nu_\relations(\Not r,x,y) \lEq \lNot\nu_\relations(r,x,y))\]
Similarly as for the logic $\Km$, 
a \emph{model} $\I$ for $\KmNot$ is a tuple $\I=(\Delta^\I,\nu_\formulae^\I,\nu_\relations^\I)$,
where $\Delta^\I$ is not empty and all the formulae of the semantic specification for $\KmNot$
are true in $\I$.

The logic $\K_m(\Not)$ is interesting %
because of the presence of
three quantifier operators. These are
necessity operator $\pBox{\alpha}$, the possibility operator
$\Not \pBox{\alpha} \Not$ and a third operator, $\pBox{\Not\alpha}\Not$,
called the sufficiency operator
sometimes referred to as the window operator.
$\nu_\formulae(\pBox{\alpha}\phi,v)$ can be read as saying $\phi$
is true in \emph{all} $\alpha$-successors, 
$\nu_\formulae(\Not\pBox{\alpha}\Not\phi,v)$
as $\phi$ is true in \emph{some} $\alpha$-successor,
and $\nu_\formulae(\pBox{\Not\alpha}\Not\phi,v)$
as $\phi$ is true in \emph{only}
$\alpha$-successors of $v$.
Following~\cite{Humberstone87}, we call $\K_m(\Not)$ the modal logic
of `some, all and only'.
$\KmNot$ is a sublogic of Boolean modal logic~\cite{GargovPassyTinchev87}
and the description logics \ALBO and \ALBOid~\cite{SchmidtTishkovsky-ALBOid-2012}.
$\KmNot$ has the finite model property~\cite{GargovPassyTinchev87}
but the tree model property fails for the logic (see, e.g., \cite{LutzSattler02}). 
The results of~\cite{LutzSattler02} imply 
that the satisfiability problem in $\KmNot$ is \ExpTime-complete.

\section{Tableau synthesis framework}\label{section:framework}
In order to synthesise a tableau calculus for a given logic $L$,
the tableau synthesis framework operates with two
languages: $\Lang{L}$, the language of specification of syntax of the logic,
and $\FO(L)$, the language of specification of semantics of the logic.
Examples of these languages for the logics $\Km$ and $\KmNot$
are defined in the previous section.

The syntax specification language $\Lang{L}$ 
is a propositional, possibly multi-sorted language.
The set of sorts of $\Lang{L}$ is denoted by $\Sorts$
and the set of the formulae of each sort $\s$ 
is denoted by $\Lang{L}^s$. 
The semantic specification language $\FO(L)$
is a multi-sorted first-order language with equality
(denoted by $\approx$).
$\FO(L)$ contains an additional \emph{domain} sort~$\D$
equipped with function and predicate symbols %
necessary for the specification of the semantics of the logic.
Expressions of $\Lang{L}$ are embedded into $\FO(L)$ as
terms of appropriate sorts
and $\FO(L)$ contains additional \emph{interpretation} symbols $\nu_\s$
for each sort $\s$ of the logic.
Depending on the sort $\s$, $\nu_\s$ can either be a functional symbol,
mapping formulae of sort $\s$ into the domain sort,
or it can be a predicate symbol of sort $(\s,\D,\ldots,\D)$.
If $\nu_\s$ is a predicate symbol then
we refer to $\nu_\s$ as `holds' or `satisfaction' predicate.

A formula $\phi$ of $\FO(L)$ is called \emph{\Lang{L}-atomic}
if it is an atomic formula of $\FO(L)$
and all occurrences of formulae of $\Lang{L}$ 
are also atomic in $\phi$.
Thus, $\nu_\s(E,\seq{t})$ is $\Lang{L}$-atomic 
only if $E$ is an atomic formula of $\Lang{L}^\s$.
For example, the formulae 
$\nu_\formulae(p,x)$ and $\nu_\relations(r,g(r,x),x)$
are $\Lang{L}(\KmNot)$-atomic, 
but the formulae $\lNot\nu_\formulae(p,x)$,
$\nu_\formulae(p\Or q,x)$, and $\nu_\relations(\Not r,g(r,x),x)$
are not. %

A semantic specification of a logic $L$
is a set $S$ of formulae of $\FO(L)$
which satisfies additional properties.
In particular, $S$ must define connectives of $L$ and contain 
only formulae of a special form  
(see \emph{normalised semantic specification} in~\cite{SchmidtTishkovsky-ASTC-lmcs-2011} for details).
An \emph{$\Lang{L}$-structure} $\I$ is a tuple
$\I\define(\Delta^\I,f^\I,\ldots,P^\I,\ldots,\{\nu_\s^\I\}_{\s\in\Sorts})$,
where $\Delta^\I$ is a non-empty set, 
$f^\I$ and $P^\I$ are interpretations of function and, respectively, predicate symbols of 
the domain sort and, for each $\s\in\Sorts$,
$\nu_\s^\I$ is an interpretation of the symbol $\nu_\s$ in $\I$.
An \emph{$L$-model} is an $\Lang{L}$-structure $\I$
such that all formulae of the semantic specification of the logic $L$ 
are true in $\I$ (for all possible interpretations of individual variables).

Within the tableau synthesis framework, 
the language $\FO(L)$ also plays the role of the tableau language.
A \emph{tableau calculus} is a set of inference rules
which have the general form 
$
\inlinetableaurule{X_0}{X_1 \tor \cdots \tor X_n},
$
where both the numerator $X_0$ and all denominators $X_i$ 
are finite sets of negated or unnegated atomic formulae 
in the language $\FO(L)$.
The formulae in the numerator are called \emph{premises}, while the
formulae in the denominators are called \emph{conclusions}.
The numerator and all the denominators are non-empty, but $n$ may be zero, in which case
the rule is a \emph{closure rule} (also written $X_0/\bot$).

A \emph{tableau derivation} or \emph{tableau} in a tableau calculus $T$ is a finitely branching,
ordered tree whose nodes are sets of formulae in $\FO(L)$.
Assuming that~$N$ is the input set of $\Lang{L}$-formulae
to be tested for satisfiability the root node of the tableau is the
set $\{\nu_\s(E,\seq{a}) \mid E \in N\cap\Lang{L}^\s, \s\in\Sorts\}$, where $\seq{a}$ denotes a 
sequence of fresh 
constant from the domain sort of an appropriate length.

Successor nodes are constructed in accordance with the inference
rules in the calculus $T$.
An inference rule $\inlinetableaurule{X_0}{X_1 \tor \cdots \tor X_n}$
is applicable to a selected formula $\phi$ in a
node of the tableau, if $\phi$, together with other
formulae in the node, are simultaneous
instantiations of formulae in $X_0$. %
Then $n$~successor nodes are created which contain the formulae of
the current node and the appropriate instances of~$X_i$.

We use the notation $T(N)$ for a (in the
limit) finished tableau built
by applying the rules of the calculus $T$ starting with the set
$N$ (of $\Lang{L}$-formulae) as input.
That is, we assume that all branches in the tableau are fully expanded and
all applicable rules have been applied in~$T(N)$.

In a tableau, a maximal path from the root node is called a
\emph{branch}.
For a branch~$\branch{B}$ of a tableau we write $\phi\in\branch{B}$
to indicate that the formula $\phi$ belongs to a node of the
branch~$\branch{B}$.
Considering any branch as a set of formulae 
it can be shown that the order of rule applications
is not essential for a tableau derivation
in the sense
that all tableau derivations started from given input $N$
contain same set of branches (as sets of formulae).
Thus, without loss of generality, we 
assume that $T(N)$ is unique.

A branch of a tableau is \emph{closed} if a closure rule
has been
applied
in this branch, otherwise the branch is called \emph{open}.
The tableau $T(N)$ is \emph{closed} if all its branches are
closed and $T(N)$ is \emph{open} otherwise.
The calculus $T$ is \emph{sound} iff 
for any (possibly infinite) set of formulae~$N$, %
$T(N)$ is open whenever $N$ is satisfiable.
$T$ is \emph{complete} iff 
$T(N)$ is closed
for any (possibly infinite) unsatisfiable set %
$N$. %

We say that a tableau calculus $T$ is \emph{constructively complete}
for a logic $L$ 
if for any open branch $\branch{B}$ in a derivation in $T$
there is an $L$-model $\IB$ such that
all the formulae in $\branch{B}$ are true in $\IB$.
Clearly, if $T$ is constructively complete for a logic $L$ then $T$ is complete for $L$. 
Following the tableau synthesis framework,
we assume that the domain of the model $\IB$
is constructed from terms of the domain sort $\D$ modulo
equalities derived in the branch $\branch{B}$.
In particular, the domain of the model is $\Delta^\IB\define\{\ecl{t}\mid\text{$t$ is a term of the domain sort and $t$ occurs in $\branch{B}$}\}$,
where $\ecl{t}\define\{t'\mid t\approx t'\in\branch{B}\}$.
We say that $\branch{B}$ is \emph{reflected} in $\IB$
iff all the formulae in $\branch{B}$ are true in $\IB$
under the valuation $t\mapsto\ecl{t}$ for each domain term $t$
(see~\cite{SchmidtTishkovsky-ASTC-lmcs-2011} for details).

Given a semantic specification for a logic $L$,
which satisfies additional conditions 
(see \emph{well-defined semantical specification} in~\cite{SchmidtTishkovsky-ASTC-lmcs-2011}),
the tableau synthesis framework 
generates a tableau calculus
sound and constructively complete for~$L$.

The tableau calculi $T_{\Km}$ and $T_{\KmNot}$ respectively 
generated in the tableau synthesis framework
from the semantic specifications for $\Km$ and $\KmNot$ 
are given in Figure~\ref{figure:Km  and KmNot tableau calculi}.
The calculus $T_\KmNot$ extends $T_\Km$ by two additional rules
for relational negation.
\begin{figure}[t]
\begin{gather*}
\intertext{Tableau rules of $T_\Km$:\vspace*{-8mm}}
\tableaurule{\nu_\formulae(\Not p, x)}{\lNot\nu_\formulae(p,x)}
\qquad\qquad
\tableaurule{\lNot\nu_\formulae(\Not p, x)}{\nu_\formulae(p,x)}
\\
\tableaurule{\nu_\formulae(p\Or q, x)}{\nu_\formulae(p,x)\tor\nu_\formulae(q,x)}
\qquad\qquad
\tableaurule{\lNot\nu_\formulae(p\Or q, x)}{\lNot\nu_\formulae(p,x)\tand\lNot\nu_\formulae(q,x)}
\\
\tableaurule{\nu_\formulae(\pBox{r}p,x)}{\lNot\nu_\relations(r,x,y)\tor\nu_\formulae(p,y)}
\qquad\qquad
\tableaurule{\lNot\nu_\formulae(\pBox{r}p, x)}{\nu_\relations(r,x,f(r,p,x))\tand\lNot\nu_\formulae(p,f(r,p,x))}
\\
\tableaurule{\nu_\formulae(p,x)\tand\lNot\nu_\formulae(p,x)}{\bot}
\qquad\qquad
\tableaurule{\nu_\relations(r,x,y)\tand\lNot\nu_\relations(r,x,y)}{\bot}
\\[-3mm]
\intertext{Additional rules of $T_{\KmNot}$:\vspace{-3mm}}
\tableaurule{\nu_\relations(\Not r,x,y)}{\lNot\nu_\relations(r,x,y)}
\qquad\qquad
\tableaurule{\lNot\nu_\relations(\Not r,x,y)}{\nu_\relations(r,x,y)}
\end{gather*}
\caption{Generated tableau calculi for $\Km$ and $\KmNot$}\label{figure:Km  and KmNot tableau calculi} 
\end{figure}
Because the semantic specifications for $\Km$ and $\KmNot$
are well-defined in the sense of~\cite{SchmidtTishkovsky-ASTC-lmcs-2011},
from Theorems~5.1 and~5.6 in~\cite{SchmidtTishkovsky-ASTC-lmcs-2011}
we immediately obtain the following result.
\begin{theorem}[Soundness and constructive completeness]%
\hspace{-1pt}The calculi~$T_\Km$ and~$T_\KmNot$ are sound and constructively complete for the logics $\Km$ and $\KmNot$. 
\end{theorem}

\section{Existing refinements}\label{section:existing refinements}
It this section we briefly recall two refinement techniques in the tableau synthesis framework~\cite{SchmidtTishkovsky-ASTC-lmcs-2011}.
The first refinement addresses the problem that,
in general,
the degree of branching of the generated rules 
is not optimal and higher than is
necessary.
The refinement reduces the number of branches of a rule by
constraining the rule with additional premises and having fewer 
conclusions. 
We refer to this refinement as \emph{rule refinement}.
Suppose~$\rho$ is a tableau rule in a sound and constructively complete
tableau calculus $T_L$ for a logic $L$.
Suppose $\rho$ has the form
$
 \rho\define\inlinetableaurule{X_0}{X_1\tor\cdots\tor X_m}.
$
Let $X_i=\{\psi_1,\ldots,\psi_k\}$ be one of the denominators of~$\rho$ for some $i\in\{1,\ldots,m\}$.
For simplicity and
without loss of generality, we assume that $i=1$.

Let the rules $\rho_j$ with $j=1,\ldots,k$ be defined by
\[
 \rho_j\define\tableaurule{X_0\cup\{\inverse\psi_j\}}{X_2\tor\cdots\tor X_m}.
\]
Each $\rho_j$ is obtained from the rule $\rho$ by removing the first
denominator $X_1$ and adding the negation of one of the formulae in $X_1$ as a premise.

We denote by $\Rfr(\rho,T_L)$ the \emph{refined tableau calculus} obtained from $T_L$
by replacing the rule $\rho$ with rules $\rho_1,\ldots,\rho_k$.
We say that $\Rfr(\rho,T_L)$ is a \emph{refinement} of $T_L$.
One can show that each $\rho_j$ is derivable~\cite{Gore-HandbookOnTableauMethods-1999} in $T_L$ and
this implies that
the calculus $\Rfr(\rho,T_L)$ is sound.
In general, $\Rfr(\rho,T_L)$ is 
neither constructively complete nor complete.
Nevertheless, the following theorem is proved in~\cite{SchmidtTishkovsky-ASTC-lmcs-2011}.

\begin{theorem}\label{theorem: tableau transformation1}
Let $T_L$ be a tableau calculus which is sound and constructively complete for 
the logic $L$.
Let $\rho$ be the rule ${X_0}/{X_1\tor\cdots\tor X_m}$ in $T_L$ and suppose
$\Rfr(\rho,T_L)$ is a refinement of $T_L$. 
Further, suppose $\branch{B}$ is an open branch in a $\Rfr(\rho,T_L)$-tableau
derivation
and for every set $Y$ of $\Lang{L}$-formulae from $\branch{B}$ the following holds.
\vspace*{-\medskipamount}
\begin{trivlist}
 \item\textbf{\upshape General rule refinement condition:}
    If all formulae in $Y$ are reflected in $\IB$
    then for any $E_1,\ldots,E_l\in Y$ and any domain terms $t_1,\ldots,t_n$
\begin{gather*}
    \begin{aligned}
        &\text{if}\ X_0(\seq{E},t_1,\ldots,t_n)\subseteq\branch{B}\ \text{and}\ 
        \IB\not\models X_1(\seq{E},\ecl{t_1},\ldots,\ecl{t_n})\\ 
        &\text{then}\ 
        X_i(\seq{E},t_1,\ldots,t_n)\subseteq\branch{B},\ \text{for some}\ i=2,\ldots,m.
    \end{aligned}%
\end{gather*}
\end{trivlist}
Then, $\branch{B}$ is reflected in $\IB$.
\end{theorem}
Assuming that $p_1,\ldots,p_l$ and $x_1,\ldots,x_n$
are respectively all the $\Lang{L}$-variables and
all the domain variables
occurring in the rule $\rho$,
$X_i(\seq{E},t_1,\ldots,t_n)$ denotes
the set of all instances
of the $\FO(L)$-formulae from $X_i$
under uniform substitution of 
$E_1,\ldots,E_l$
and 
$t_1,\ldots,t_n$
into 
$p_1,\ldots, p_l$ and $x_1,\ldots,x_n$ respectively.

The general rule refinement condition states
that if there is not enough information in the branch $\branch{B}$ 
to derive the formulae of $X_1(\seq{E},t_1,\ldots,t_n)$
in the model constructed from $\branch{B}$
then at least one of the other denominators of the rule is explicitly
contained in the branch $\branch{B}$.

The general rule refinement condition corresponds to the condition~($\ddagger$)
in~\cite{SchmidtTishkovsky-ASTC-lmcs-2011}
which is stronger than condition~($\dagger$)
in~\cite[Theorem~6.1]{SchmidtTishkovsky-ASTC-lmcs-2011}
but is enough for the purposes of this paper.
A consequence of the theorem is the following.
\begin{corollary}\label{corollary: refinement preserves constructive completeness}
If the condition of Theorem~\ref{theorem: tableau transformation1}
holds for every open branch $\branch{B}$ of any fully expanded $\Rfr(\rho,T_L)$-tableau
then the refined calculus $\Rfr(\rho,T_L)$ is constructively complete for the logic $L$.
\end{corollary}

Generalising this refinement to turning more than one denominator into
premises is not difficult. 

The second refinement in the framework allows to internalise~$\nu_\s$
and the domain sort symbols inside the language of the logic
if there are appropriate constructs in $\Lang{L}$ with
the same semantics. 
In this case each atomic formula $\nu_\s(E,\seq{a})$
in the tableau calculus for $L$ is 
replaced by a suitable formula of the logic
and, then, all syntactically redundant rules
are removed from the transformed calculus.
This refinement simplifies the tableau language
and, in many cases, reduces the number
of the rules in the tableau calculus.
We refer to this refinement as the \emph{internalisation refinement}.

The intended way to apply these two refinement is
in the order of their description here.
Usually, this order is also the easiest way 
for applying the refinements and  
produces the best possible improvement 
of the generated calculus.

\section{Using existing refinements}\label{section:examples of refinement}
The box decomposition rule\vspace*{-3mm}
\[\tableaurule{\nu_\formulae(\pBox{r}p,x)}{\lNot\nu_\relations(r,x,y)\tor\nu_\formulae(p,y)}[\textsf{box}]\label{rule:box}\]
in the calculus $T_\Km$
can be refined to the usual box rule 
\[\tableaurule{\nu_\formulae(\pBox{r}p,x)\tand\nu_\relations(r,x,y)}{\nu_\formulae(p,y)}[$\Box$]\label{rule:box+}\]
preserving constructive completeness of the calculus.
It can be proved directly that
the generic refinement condition 
is true for this rule in any branch of $\Rfr(\ref{rule:box},T_\Km)$-derivation,
and, thus, by Corollary~\ref{corollary: refinement preserves constructive completeness}
the calculus $\Rfr(\ref{rule:box},T_\Km)$ is still constructively complete.
(We notice that constructive completeness of $\Rfr(\ref{rule:box},T_\Km)$ also follows from 
Corollary~\ref{corollary: atomic refinement preserves constructive completeness} in Section~\ref{section:atomic rule refinement}
because any instantiation of $\nu_\relations(r,x,y)$ in the language of $\Km$
is an $\Lang{L}(\Km)$-atomic formula.)
\begin{theorem}\label{theorem:completeness of Km}
The tableau calculus $\Rfr(\ref{rule:box},T_{\Km})$ 
is sound and constructively complete for the logic $\Km$.
\end{theorem}

However, %
none of the rules of the tableau calculus for $\KmNot$ from Figure~\ref{figure:Km  and KmNot tableau calculi}
are refinable. 
In particular, the \ref{rule:box}~rule cannot be refined to the \ref{rule:box+} rule
without loosing constructive completeness.
Take for instance the set of formulae $\{\nu_\formulae(\pBox{\Not\Not r}p,a), \nu_\relations(r,a,b), \lNot\nu_\formulae(p,b)\}$.
It is $\KmNot$-unsatisfiable but none of the rules of the refined calculus $\Rfr(\ref{rule:box},T_{\KmNot})$
are applicable to the set.

Nevertheless, using a small transformation trick with the semantic specification
we can obtain a tableau calculus where this refinement is possible.
We observe that the following statement is derivable from
the semantic specification of $\KmNot$.
\[\forall x\; (\nu_\formulae(\pBox{\Not r}p, x)\lImp \forall y\;(\lNot \nu_\relations(r,x,y)\lImp\nu_\formulae(p,y)))\]
This means that it can be added to the semantic specification of 
$\KmNot$ without changing the class of models of the logic.
We denote the tableau calculus generated from the semantic specification 
extended with this statement
by $T_{\KmNot}^+$.
$T_{\KmNot}^+$ consists of the rules listed in Figure~\ref{figure:Km  and KmNot tableau calculi}
and %
the following additional rule.
\[
 \tableaurule{\nu_\formulae(\pBox{\Not r}p,x)}{\nu_\relations(r,x,y)\tor\nu_\formulae(p,y)}[$\pBox{\Not}$]\label{rule:box not}
\]
It is possible to check that
the well-definedness conditions from~\cite{SchmidtTishkovsky-ASTC-lmcs-2011}
are satisfied
for the extended semantic specification of $\KmNot$.
Therefore, by results in the tableau synthesis framework,
$T_{\KmNot}^+$ is sound and constructively complete for~$\KmNot$.

The general rule refinement condition is now satisfied
for the calculus obtained
from $T_{\KmNot}^+$ by refinement of the \ref{rule:box} rule and, thus,
the following theorem is a consequence of Corollary~\ref{corollary: refinement preserves constructive completeness}.
\begin{theorem}\label{theorem:completeness of KmNot}
The tableau calculus %
$\Rfr(\ref{rule:box},T_{\KmNot}^+)$ is sound and constructively complete for the logic $\KmNot$.
\end{theorem}

The internalisation refinement 
is possible for the calculi in accordance\linebreak with~\cite{SchmidtTishkovsky-ASTC-lmcs-2011} 
if nominals and the `satisfaction' operator $\at{}$
of hybrid logic~\cite{Blackburn-WHL-1998}
are introduced to the tableau languages of $\Km$ and $\KmNot$.
More precisely, 
every formula $\nu_\formulae(\phi,a)$ is replaced by
the formula $\at{a}\phi$,
and every $\nu_\relations(\alpha,a,b)$ is replaced by
the formula $\at{a}\Not\pBox{\alpha}\Not b$.
In this case the results of the refinement are \emph{labelled tableau calculi},
which are also sound and constructively complete for $\Km$ and $\KmNot$.
Their rules are listed in Figure~\ref{figure:Km and KmNot refined tableau}.
We denote these calculi by $T_\Km^r$ and $T_\KmNot^r$ respectively.
\begin{figure}[t]
\begin{gather*}
\intertext{Refined tableau rules for $\Km$:\vspace*{-3mm}}
\tableaurule{\at{i} p\tand\at{i}\Not p}{\bot}
\qquad\qquad
\tableaurule{\at{i}\Not\Not p}{\at{i}p}
\\
\tableaurule{\at{i}(p\Or q)}{\at{i}p\tor\at{i}q}
\qquad\qquad
\tableaurule{\at{i}\Not(p\Or q)}{\at{i}\Not p\tand\at{i}\Not q}
\\
\tableaurule{\at{i}\pBox{r}p\tand\at{i}\Not\pBox{r}\Not j}{\at{j}p}
\qquad\qquad
\tableaurule{\at{i}\Not\pBox{r}p}{\at{i}\Not\pBox{r}f(r,p,i)\tand\at{f(r,p,i)}\Not p}
\\[-3mm]
\intertext{Additional refined rules for $\KmNot$:\vspace*{-3mm}}
\tableaurule{\at{i}\Not\pBox{\Not r}\Not j}{\at{i}\pBox{r}\Not j}
\qquad\qquad
\tableaurule{\at{i}\pBox{\Not r}\Not j}{\at{i}\Not\pBox{r}\Not j}
\qquad\qquad
\tableaurule{\at{i}\pBox{\Not r}p}{\at{i}\Not\pBox{r}\Not j\tor\at{j}p}
\end{gather*}
\caption{Refined tableau calculi $T_\Km^r$ %
and $T_\KmNot^r$%
.}\label{figure:Km and KmNot refined tableau} 
\end{figure}

\section{Atomic rule refinement}\label{section:atomic rule refinement}
In this section we introduce the technique of 
\emph{atomic rule refinement}.
Under this refinement,
all conclusions of a rule 
which are moved upward %
are negated \Lang{L}-atomic formulae of the language \FO(L).
More precisely, in the notation and with the assumptions of Theorem~\ref{theorem: tableau transformation1},
the following result holds.
\begin{theorem}\label{theorem: tableau transformation: atoms}
Assume that
for an open branch $\branch{B}$ of the refined tableau $\Rf(\rho,T_L)$
and 
 for every set $Y$ of $\Lang{L}$-formulae from $\branch{B}$ the following holds.
\vspace*{-\medskipamount}
\begin{trivlist}
 \item\textbf{\upshape Atomic rule refinement condition:}
    If all formulae in $Y$ are reflected in $\IB$
    then for any $E_1,\ldots,E_l\in Y$ and any domain terms $t_1,\ldots,t_n$,
\begin{align*}
    \begin{aligned}
        & X_0(\seq{E},t_1,\ldots,t_n)\subseteq\branch{B}\ \text{implies that}\ \\
        & X_1(\seq{E},t_1,\ldots,t_n)=\{\Not\xi_1,\ldots,\Not\xi_k\}\ \text{and all $\xi_1,\ldots,\xi_k$ are \emph{\Lang{L}-atomic}}.
    \end{aligned}
\end{align*}
\end{trivlist}
Then, $\branch{B}$ is reflected in $\IB$.
\end{theorem}

Unlike the general rule refinement condition,
the atomic rule refinement condition
is purely syntactic and, thus,
can be automatically checked against each given open branch $\branch{B}$.
However, even if all the formulae from $X_1$ are negated $\Lang{L}$-atomic
their instantiation within a branch of a tableau derivation
can, in general, produce a formula which is not a negated $\Lang{L}$-atom.
Therefore, similar to Corollary~\ref{corollary: refinement preserves constructive completeness},
by Theorem~\ref{theorem: tableau transformation: atoms},
in order to preserve constructive completeness of the calculus
under atomic rule refinement we need to make sure that
the atomic rule refinement condition
holds for every branch of any derivation in the refined calculus.
\begin{corollary}\label{corollary: atomic refinement preserves constructive completeness}
If the assumptions and condition of Theorem~\ref{theorem: tableau transformation: atoms}
holds for every open branch $\branch{B}$ of any fully expanded $\Rfr(\rho,T_L)$-tableau
then the refined calculus $\Rfr(\rho,T_L)$ is constructively complete for the logic $L$.
\end{corollary}

\section{Atomic rule refinement for frame conditions}\label{section:refinement for frame conditions}
In this and the following section we consider two important cases
in which Corollary~\ref{corollary: atomic refinement preserves constructive completeness} holds. 

The first case is important because
it allows to automatically refine 
tableau rules generated from 
frame conditions of modal logics.

Consider the axiom of irreflexivity of $\Km$-relations
introduced in Section~\ref{section:logics}: $\forall x\,\lNot\nu_\relations(r,x,x)$.
The rule generated from this property is
$\tableaurule{}{\lNot\nu_\relations(r,x,x)}\label{rule:irreflexivity}$.
We claim that this rule can be refined
to the following closure rule
\[\tableaurule{\nu_\relations(r,x,x)}{\bot}[irr]\label{rule:irreflexivity+}.\]
Because the language of $\Km$ contains
only atomic relations $a_1,\ldots,a_m$
and no relational operators,
any instantiation of $r$ (and variable $x$)
in $\nu_\relations(r,x,x)$ produces 
only $\Lang{L}(\Km)$-atomic formulae
of the form $\nu_\relations(a_i,t,t)$
(where $t$ is a term of the domain sort).
Therefore, the atomic rule refinement condition is true
for any branch of any tableau derivation
in the calculus $\Rfr(\ref{rule:box},T_\Km)$ extended with the \ref{rule:irreflexivity+} rule.
Thus, by Corollary~\ref{corollary: atomic refinement preserves constructive completeness},
the calculus $\Rfr(\ref{rule:box},T_\Km)$ extended with the \ref{rule:irreflexivity+} rule
is sound and constructively complete for the logic $\Km$ with irreflexive relations.
Applying internalisation refinement
we obtain the following theorem for the labelled tableau calculus.
\begin{theorem}
$T_\Km^r$ extended with the rule $\inlinetableaurule{\at{i}\Not\pBox{r}\Not i}{\bot}$
is sound and constructively complete for $\Km$ with irreflexive relations.
\end{theorem}

For another example consider the frame condition stating
existence of an immediate predecessor of any element of a model
in Section~\ref{section:logics}.
We reduce it to a form which is acceptable in the tableau synthesis framework.
Let $g$ be a new Skolem function which depends on the two arguments of the sort of relations
and the domain sort.
We remove the existential quantifier from the frame condition
and decompose the result into three formulae:
\begin{gather*}
\forall x\;\nu_\relations(r,g(r,x),x),\qquad\forall x\; (x\not\approx g(r,x)),\\%
\forall x\forall z\;\bigl((\nu_\relations(r,g(r,x),z)\lAnd\nu_\relations(r,z,x))\lImp(g(r,x)\approx z\lOr z\approx x)\bigr).
\end{gather*}
From these formulae three rules are generated:
\begin{gather*}
\tableaurule{}{\nu_\relations(r,g(r,x),x)},\qquad\tableaurule{}{x\not\approx g(r,x)},\\%
\tableaurule{}{\lNot\nu_\relations(r,g(r,x),z)\tor\lNot\nu_\relations(r,z,x)\tor g(r,x)\approx z\tor z\approx x}.
\end{gather*}
The atomic rule refinement is not applicable to the first rule since
the conclusion is not negated.
Consider the second and third rules.
Applying the same argument as for the rule generated from the irreflexivity axiom
we find that any instantiation of 
$x\approx g(r,x)$, 
$\nu_\relations(r,g(r,x),z)$,
and $\nu_\relations(r,z,x)$
within the language $\FO(\K_m)$ cannot
produce a formula which is not $\Lang{L}(\Km)$-atomic.
Hence, the atomic rule refinement condition holds for these rules
in any branch of any tableau derivation constructed within the language $\FO(\K_m)$.
Therefore, refining the second rule once and the third rule twice
the rules
\[\tableaurule{x\approx g(r,x)}{\bot}\qquad\text{and}\qquad%
\tableaurule{\nu_\relations(r,g(r,x),z)\tand\nu_\relations(r,z,x)}{g(r,x)\approx z\tor z\approx x}\]
are obtained.
By Corollary~\ref{corollary: atomic refinement preserves constructive completeness},
constructive completeness 
of any tableau calculus in the language $\FO(\K_m)$
is preserved under these refinements.
Internalising $\FO(\K_m)$ in the hybrid logic extension of $\Km$
and introducing a new function $g$ which depends on two arguments
of the relational sort and, respectively, the sort of nominals
(see~\cite{SchmidtTishkovsky-ASTC-lmcs-2011} for details)
we, in particular, obtain the following theorem.
\begin{theorem}
$T_\Km^r$ extended with the rules
\[%
\tableaurule{}{\at{g(r,i)}\Not\pBox{r}\Not i},%
\qquad
\tableaurule{\at{i}g(r,i)}{\bot}%
\qquad\text{and}\qquad%
\tableaurule{\at{g(r,i)}\Not\pBox{r}\Not j\tand\at{j}\Not\pBox{r}\Not i}{\at{g(r,i)}j\tor\at{j}i}\]
is sound and constructively complete for $\Km$ 
over the class of models 
satisfying the frame condition of existence of an immediate predecessor.
\end{theorem}

\section{Hypertableau}\label{section:hypertableau}
Let the given logic $L$ have disjunction-like connectives $\Or$
and negation-like connectives $\Not$
for some sort $\s$ of the logic. 
Assume $T_L$ is a tableau calculus sound and constructively complete for 
$L$ and contains the rules 
\[
 \tableaurule{\nu_\s(\Not p,\seq{x})}{\lNot\nu_\s(p,\seq{x})}\quad\text{and}\quad\tableaurule{\nu_\s(p\Or q,\seq{x})}{\nu_\s(p,\seq{x})\tor\nu_\s(q,\seq{x})},
\]
which are the usual rules for disjunction and negation.
We transform the calculus $T_L$ into a new calculus $T_L^{\textsf{hyp}}$ in three 
steps.
For simplicity we assume that disjunction in %
$L$ is associative and commutative
with respect to satisfiability, that is, the following statements are derivable from the semantic specification~of~$L$:
\begin{align*}
 \nu_\s(p\Or q,\seq{x})&\lEq\nu_\s(q\Or p,\seq{x}),\\
 \nu_\s((p\Or q)\Or r,\seq{x})&\lEq\nu_\s(p\Or(q\Or r),\seq{x}).
\end{align*}
This assumption is not essential for the transformation but
allows to simplify disjunctions
and avoid a combinatorial blow-up.

In the first step of the transformation, the usual disjunction rule \linebreak
$\inlinetableaurule{\nu_\s(p\Or q,\seq{x})}{\nu_\s(p,\seq{x})\tor\nu_\s(q,\seq{x})}$
of $T_L$ is replaced by
the set of the %
rules (for $k>1$):
\[
\tableaurule{\nu_\s(p_1\Or\cdots\Or p_k,\seq{x})}{\nu_\s(p_1,\seq{x})\tor\cdots\tor\nu_\s(p_k,\seq{x})}[split${}_k$]\label{rule:clause}.
\]
We denote by $T_L^c$ a tableau calculus obtained from $T_L$
by replacing the usual disjunction rule by the rules~\ref{rule:clause}. 
The \ref{rule:clause} rules and the usual disjunction rule are derivable from each other.
Therefore, the transformed $T_L^c$ calculus is sound and constructively complete.

For the second step consider the following rules (for $m+n>1$).
\begin{gather*}
\tableaurule{\nu_\s(\Not p_1\Or\cdots\Or\Not p_m\Or q_1\Or\cdots\Or q_n,\seq{x})}%
{\lNot\nu_\s(p_1,\seq{x})\tor\cdots\tor\lNot\nu_\s(p_m,\seq{x})\tor\nu_\s(q_1,\seq{x})\tor\cdots\tor\nu_\s(q_n,\seq{x})}[split${}^+_{mn}$]\label{rule:clause'}\\
(\text{only atomic substitutions are allowed into $p_1,\ldots,p_m$})
\end{gather*}
That is,
the rules are applicable only to formulae of the shape
$\nu_\s(\Not E_1\Or\cdots\Or\Not E_m\Or F_1\Or\cdots\Or F_n,\seq{x})$, where
all $E_1,\ldots,E_m$ are \emph{atomic} formulae of the logic~$L$.
We also implicitly assume that all $F_1,\ldots,F_n$ are not negated atomic formulae of $L$.
Let $T_L^{c+}$ be a tableau calculus obtained from $T_L^c$
by replacing the rules \ref{rule:clause} by the rules \ref{rule:clause'}.
The rules \ref{rule:clause} and the rules \ref{rule:clause'}
are derivable from each other and, thus, the following theorem holds.
\begin{theorem}\label{theorem:T_L^{c+}}
 $T_L^{c+}$ is sound and constructively complete for the logic $L$.
\end{theorem}

In the final step, we refine the rules obtained in the previous step
to the set of rules ($m+n>1$)
which are %
hypertableau-like rules.
\begin{gather*}
\tableaurule{\nu_\s(\Not p_1\Or\cdots\Or\Not p_m\Or q_1\Or\cdots\Or q_n,\seq{x})\tand \nu_\s(p_1,\seq{x})\tand\cdots\tand\nu_\s(p_m,\seq{x})}%
{\nu_\s(q_1,\seq{x})\tor\cdots\tor\nu_\s(q_n,\seq{x})}[hyp${}_{mn}$]\label{rule:hypertableau}\\
(\text{only atomic substitutions are allowed into $p_1,\ldots,p_m$})
\end{gather*}
Similarly to the rules in the previous step, 
an application of the rule~\ref{rule:hypertableau} is allowed
only to formulae of the shape
$\nu_\s(\Not E_1\Or\cdots\Or\Not E_m\Or F_1\Or\cdots\Or F_n)$, where
all $E_1,\ldots,E_m$ are atomic formulae 
and $F_1,\ldots,F_n$ are all not negated atomic formulae of the logic $L$.
Notice that in the case of $n=0$ the rules~\ref{rule:hypertableau} are atomic closure rules.

Let $T_L^{\mathsf{hyp}}$ be the calculus
obtained from $T_L^c$ by adding the \ref{rule:hypertableau} rules.
By Corollary~\ref{corollary: atomic refinement preserves constructive completeness} and Theorem~\ref{theorem:T_L^{c+}}
we obtain constructive completeness of $T_L^{\textsf{hyp}}$.
\begin{theorem}\label{theorem:hypertableau}
$T_L^{\textsf{hyp}}$ is sound and constructively complete for the logic $L$. 
\end{theorem}
Thus, for any (propositional) logic $L$ with disjunction and negation connectives and
any sound and constructive complete calculus for $L$
with the usual disjunction and negation rules,
it is possible to devise
a hypertableau-like calculus that is sound and constructively complete for the logic $L$.

Derivations in $T_L^{\textsf{hyp}}$
can be done more efficiently
if the given logic $L$ has additional properties.
We have already assumed associativity and commutativity of disjunction.
Suppose now that satisfiability of formulae
in a large subset of the language of $L$ 
is reducible to satisfiability
of a set of clauses of formulae:
\[
 \nu_\s(E,\seq{x})\lEq\bigAnd_{i=1}^{I}\nu_{\s_{ij}}(\bigOr_{j=1}^{J_i} E_{ij},\seq{x}).
\]
Thus, every formula $E$ has an equi-satisfiable clausal
representation as set of clauses $C_1,\ldots,C_I$, where
$C_i = E_{i1}\Or\cdots\Or E_{iJ_i}$ for each $i=1,\ldots,I$.
Since disjunction is associative and commutative,
we can assume that, in every clause,
all negated atomic formulae (negative literals) of the logic
appear before all other formulae.
Let $\mathcal{A}$ be the reduction algorithm
which transform any formula $E$ into such equi-satisfiable 
clausal normal form.

The case becomes interesting 
when for many formulae of the logic
their clausal normal form has clauses with negated atomic formulae.
This assumption implies that the \ref{rule:hypertableau} rules
with $m>0$ are applied on average often in derivations in $T_L^{\textsf{hyp}}$.
Since the \ref{rule:hypertableau} rules with $m>0$ 
create less branching points in derivations than 
the \ref{rule:hypertableau} rule with $m=0$,
derivations in $T_L^{\textsf{hyp}}$ contain
less branches and, therefore, is more efficient.

We notice that the conclusions of the \ref{rule:hypertableau} rules
are allowed to contain non-atomic $\Lang{L}$-formulae
which have to be decomposed further by other rules of the calculus.
For the conclusions of other rules, we have two alternatives.
One is to use the rules of the tableau calculus to decompose
their formulae up to atomic components.
The other alternative is to apply the clausification algorithm
$\mathcal{A}$ to every new conclusion of 
any rule which is different from the \ref{rule:hypertableau} rules.
The first alternative uses the power of the original calculus $T_L$
and the second one uses the power of the clausification algorithm $\mathcal{A}$.
For efficiency of algorithms based on the tableau calculus $T_L^{\textsf{hyp}}$, 
these two alternatives have to be well-balanced
depending on the complexity of the algorithm $\mathcal{A}$ and how efficiently it 
is implemented.

The logic $\KmNot$ supports a Boolean disjunction and negation on the sort of formulae.
Therefore it is possible to devise a hypertableau calculus for $\KmNot$.
There is an efficient clausification algorithm for Boolean part which runs in polynomial time 
on the length of the input~\cite{NonnengartWeidenbach01}. 
Thus, we assume that every conclusion of a rule is immediately transformed into a set of clauses.
This allows to omit all the rules for Boolean connectives except the hypertableau rules.    
The hypertableau calculus for $\KmNot$ in a form of a labelled calculus 
is presented in Figure~\ref{figure:KmNot hypertableau}.
By Theorem~\ref{theorem:hypertableau}, this calculus is sound and constructively complete for~$\KmNot$.

\begin{figure}[t]
\begin{gather*}
\tableaurule{\at{i}\Not p_1\Or\cdots\Or\Not p_m\Or q_1\Or\cdots\Or q_n\tand \at{i}p_1\tand\ldots\tand\at{i}p_m}%
{\at{i}q_1\tor\cdots\tor\at{i}q_n}\left(\genfrac{}{}{0pt}{0}{m+n>1}{p_1,\ldots,p_m\text{ are atomic}}\right)
\\
\tableaurule{\at{i} p\tand\at{i}\Not p}{\bot}
\qquad\qquad
\tableaurule{\at{i}\pBox{r}p\tand\at{i}\Not\pBox{r}\Not j}{\at{j}p}
\qquad\qquad
\tableaurule{\at{i}\Not\pBox{r}p}{\at{i}\Not\pBox{r}f(r,p,i)\tand\at{f(r,p,i)}\Not p}
\\
\tableaurule{\at{i}\Not\pBox{\Not r}\Not j}{\at{i}\pBox{r}\Not j}
\qquad\qquad
\tableaurule{\at{i}\pBox{\Not r}\Not j}{\at{i}\Not\pBox{r}\Not j}
\qquad\qquad
\tableaurule{\at{i}\pBox{\Not r}p}{\at{i}\Not\pBox{r}\Not j\tor\at{j}p}
\end{gather*}
\caption{Labelled hypertableau calculus for $\KmNot$}\label{figure:KmNot hypertableau} 
\end{figure}

\section{Concluding remarks}\label{section:discussion}
The paper is an investigation of refinement techniques 
of tableau calculi developed within the tableau synthesis framework.
Rule refinement reduces the number of branches of rules and, therefore,
tableau algorithms based on refined tableau calculi
run more efficiently comparing with the algorithms based on the original calculi.
Furthermore, the refinement provides an incremental method of improving and optimising 
sound and constructive complete tableau calculi.
The most generic condition 
which ensures that constructive completeness is preserved under rule refinement
is second-order, and, thus, is difficult to check.
In contrast,
the condition for atomic rule refinement 
presented in this paper is purely syntactic and, thus,
can be easily verified.
It turns out that this kind of refinement can be applied 
in many cases of tableau calculi developed for various logics.
The refinement works for rules reflecting frame conditions
of modal logics and declarations of role properties in description logics,
and allows to develop hypertableau-like calculi
for logics with disjunction and negation.

As case studies we considered the logic $\Km$ with frame conditions and 
the logic  $\KmNot$ of `some', `all' and `only'.
We showed that the tableau calculus for $\KmNot$
generated by the tableau synthesis framework
can be made refinable by using a trick of extending semantic specification of $\KmNot$
by a new statement derivable in the original specification.
In this case, we proved that the general second-order rule refinement condition becomes true.
On the basis of the refined tableau calculus
we developed a hypertableau calculus for $\KmNot$
applying the atomic rule refinement to the calculus.

The tableau calculus of \ALBOid
from~\cite{SchmidtTishkovsky-ALBOid-2012} can be used for deciding
$\KmNot$.
\ALBOid is an extension of the description logic \ALC 
with individuals, the inverse role operator,
Boolean operators on roles and the identity role.
Although not developed with the techniques described in this paper, 
we remark the
tableau calculus of \ALBOid can be obtained by altering the semantic
specification similar as described for $\KmNot$ in this paper.

We observe that the original rule is derivable from the rules
obtained from it by the rule refinement method
if the calculus contains the analytic cut rule~\cite{DAgostinoMondadori-TC+-1994}.
Thus, this refinement preserves constructive completeness in the presence of the analytic cut rule.
Therefore, KE tableau calculi can be systematically
defined using refinement 
from calculi generated by the framework.

For simplicity of presentation 
we omitted explicit equality reasoning
from the presented tableau calculi.
However, it must be noted that if
the calculus is able to derive an equality formula
then some form of the equality reasoning must be  
performed within tableau derivations
in order to keep completeness of the calculus.
This can be done either by a special group of tableau equality rules~\cite{SchmidtTishkovsky-ASTC-lmcs-2011}
or by means of ordered rewriting as it is implemented in \mettelII prover generator~\cite{TishkovskySchmidtKhodadadi-MetTeL2-2012a}.

For future work, it is of interest
to implement the considered types of tableau calculi
for different logics and compare their performance.
With the \mettelII prover generator~\cite{TishkovskySchmidtKhodadadi-MetTeL2-2012a}, this task
should be feasible but requires additional implementation efforts.
Connections of the proposed hypertableau method
and the hypertableau calculi of~\cite{BaumgartnerFurbachNiemela96,MotikShearerHorrocks-HRDL-2009}
is also a promising direction of research.

The tableau refinement methods presented in this paper as an extension
of~\cite{SchmidtTishkovsky-ASTC-lmcs-2011}
gives a novel view on existing tableau calculi and
makes development of new tableau calculi easy and accessible.

\newpage
\appendix

\section*{Proofs of theorems and statements}

For better understanding of the proofs in this section
we give a detailed formal definition of the notion of \emph{constructive
completeness} of a tableau calculus~\cite{SchmidtTishkovsky-ASTC-lmcs-2011}.
Let $\branch{B}$ denote an arbitrary open branch in a $T$-tableau
derivation.
We define an $\Lang{L}$-structure $\IB$ as follows.
Let the relation $\simB$ is defined by %
\[t\simB t'\defiff t\approx t'\in\branch{B},\]
for any ground terms $t$ and $t'$ of the domain sort $\D$ in~$\branch{B}$.
Let $\ecl{t}\define\{t'\mid t\simB t'\}$ be the equivalence class of an element $t$.
The presence of special equality rules %
ensures that~$\simB$ is a congruence
relation on all domain ground terms in~$\branch{B}$~\cite{SchmidtTishkovsky-ASTC-lmcs-2011}.
Then the domain of $\IB$ is defined as $\Delta^\IB\define\{\ecl{t}\mid t\text{ occurs in }\branch{B}\}$.
Interpretations of predicate symbols in $\IB$ are defined 
by induction on length of formulae of $\Lang{L}$ as follows: 
\begin{itemize}
 \item 
For every $n$-ary constant predicate symbol $P$,
        \[P^\IB\define\{(\ecl{t_1},\ldots,\ecl{t_n})\mid P(t_1,\ldots,t_n)\in\branch{B}\}.\]
 \item For every $\s\in\Sorts$ and $n=\ar(\s)$ 
       \begin{itemize}
	\item if $n=0$ then $\nu_\s^\IB(t)\define\ecl{\nu_\s(t)}$
for every term $t$. %
	\item if $n>0$ then
       the interpretation $\nu_\s^\IB$ is defined as
       the smallest subset of
       $\Lang{L}^\s\times(\Delta^\IB)^n$
       satisfying both the following, %
       for every variable or constant $p$ of the sort $\s$, 
       every connective $\sigma$, and any formulae $E_1,\ldots,E_m$:
\begin{align*}
       (p,\ecl{t_1},\ldots,\ecl{t_n})\in\nu_\s^\IB & \iff%
                    \nu_\s(p,t_1,\ldots,t_n)\in\branch{B},
\\
       (\sigma(\seq{E}),\ecl{t_1},\ldots,\ecl{t_n})\in\nu_\s^\IB
       & \iff \IB\models\phi^\sigma(\seq{E},\ecl{t_1},\ldots,\ecl{t_n}).
\end{align*}
      Recall that $\phi^\sigma$ denotes an \Lang{L}-open formula which defines the connective~$\sigma$.
      \end{itemize}
\end{itemize}

We say a model~$\IB$ %
\emph{reflects} a formula $E$ of the sort~$\s$
occurring in
a branch $\branch{B}$
iff for $n=\ar(\s)$ and
for all
ground terms $t_1,\ldots,t_n$ we have that
\begin{trivlist}
       \item
           $(E,\seq{\ecl{t}})\in\nu_\s^\I$ whenever
           $\nu_\s(E,\seq{t})\in\branch{B}$, and
           $(E,\seq{\ecl{t}})\notin\nu_\s^\I$ whenever $\Not\nu_\s(E,\seq{t})\in\branch{B}$.
\end{trivlist}
Similarly, $\I$ \emph{reflects} predicate constant $P$ from $\branch{B}$
iff
for all
ground terms $t_1,\ldots,t_n$ we have that
\begin{trivlist}
 \item $(\seq{\ecl{t}})\in P^\I$ whenever $P(\seq{t})\in\branch{B}$,
            and 
$(\seq{\ecl{t}})\notin P^\I$ whenever $\Not P(\seq{t})\in\branch{B}$.
\end{trivlist}
A model $\IB$ \emph{reflects} branch $\branch{B}$ %
if $\IB$ reflects all predicate constants and formulae occurring in $\branch{B}$.

A tableau calculus $T$ is said to be \emph{constructively complete} (for a logic $L$)
iff for any given set of formulae $N$, 
if $\branch{B}$ is an open branch in a tableau derivation~$T(N)$
then $\IB$ is an $L$-model which reflects $\branch{B}$.
It is clear that if $T$ is constructively complete for $L$ then $T$ is complete for $L$.

\begin{reftheorem}{theorem:completeness of Km}
The tableau calculus $\Rfr(\ref{rule:box},T_{\Km})$ 
is sound and constructively complete for the logic $\Km$.
\end{reftheorem}
\begin{proof}
We prove the general rule refinement condition of Theorem~\ref{theorem: tableau transformation1} holds
for any open branch $\branch{B}$ of $\Rfr(\ref{rule:box},T_{\Km})$. 
The result is then a consequence of Corollary~\ref{corollary: refinement preserves constructive completeness}.
Let $\nu_\formulae(\pBox{a_i}\phi,t)$ be in arbitrary open branch $\branch{B}$
of a derivation in the refined tableau calculus $\Rfr(\ref{rule:box},T_{\Km})$
and $\IB\not\models\lNot\nu_\relations(a_i,t,t')$.
Therefore, $\IB\models\nu_\relations(a_i,t,t')$.
By the definition of $\IB$, this means that
$\nu_\relations(a_i,t,t')\in\branch{B}$. 
This implies that
the refined rule \ref{rule:box+} has been applied to $\nu_\formulae(\pBox{a_i}\phi,t)$
and $\nu_\relations(a_i,t,t')$ in $\branch{B}$
and, consequently, $\nu_\formulae(\phi,t')$ is in $\branch{B}$.
\end{proof}

\begin{reftheorem}{theorem:completeness of KmNot}
The tableau calculus %
$\Rfr(\ref{rule:box},T_{\K_m(\Not)}^+)$ is sound and constructively complete for the logic $\KmNot$.
\end{reftheorem}
\begin{proof}
We prove the general rule refinement condition of Theorem~\ref{theorem: tableau transformation1} holds
for any open branch $\branch{B}$ of $\Rfr(\ref{rule:box},T_{\K_m(\Not)}^+)$. 
The result is then a consequence of Corollary~\ref{corollary: refinement preserves constructive completeness}.
Let $\nu_\formulae(\pBox{\alpha}\phi,t)$ be in arbitrary open branch $\branch{B}$
of a derivation in the refined tableau calculus $\Rfr(\ref{rule:box},T_{\K_m(\Not)}^+)$
and $\IB\not\models\lNot\nu_\relations(\alpha,t,t')$.
Therefore, $\IB\models\nu_\relations(\alpha,t,t')$.

If $\alpha$ is an atomic relation then, because $\IB\models\nu_\relations(\alpha,t,t')$,
we have that $\nu_\relations(\alpha,t,t')$ is in $\branch{B}$. 
Therefore, the refined rule \ref{rule:box+} has been applied to $\nu_\formulae(\pBox{\alpha}\phi,t)$
and $\nu_\relations(\alpha,t,t')$ in $\branch{B}$.
As a consequence, $\nu_\formulae(\phi,t')$ is in $\branch{B}$.

If $\alpha$ is not atomic then
$\alpha=\Not\alpha'$.
By induction on the length of $\alpha$, we prove that 
$\IB\models\nu_\relations(\alpha,t,t')$ implies
$\nu_\relations(\alpha',t,t')\not\in\branch{B}$.
If $\alpha'$ is atomic the case follows from the definition of $\IB$.
If $\alpha'$ is not atomic then
we have $\alpha'=\Not\alpha''$.
Thus, $\nu_\relations(\alpha',t,t')\in\branch{B}$ implies that
$\lNot\nu_\relations(\alpha'',t,t')\in\branch{B}$.
On the other hand, $\IB\models\nu_\relations(\alpha,t,t')$ if and only if 
$\IB\models\nu_\relations(\alpha'',t,t')$.
If $\alpha''$ is atomic then $\nu_\relations(\alpha'',t,t')\in\branch{B}$
by the definition of $\IB$. 
Thus, because $\branch{B}$ is open, $\lNot\nu_\relations(\alpha'',t,t')\not\in\branch{B}$.
This implies that $\nu_\relations(\alpha',t,t')\not\in\branch{B}$.
If $\alpha''=\Not\alpha'''$, then 
by the induction hypothesis we have 
$\nu_\relations(\alpha''',t,t')\not\in\branch{B}$.
Thus, $\lNot\nu_\relations(\Not\alpha''',t,t')\not\in\branch{B}$
and, consequently, $\nu_\relations(\alpha',t,t')\not\in\branch{B}$
by the rules of the calculus.
Finally, the rule \ref{rule:box not} 
was applied to $\nu_\formulae(\pBox{\Not\alpha'}\phi,t)$
in $\branch{B}$ and, hence, %
$\branch{B}$ contains either $\nu_\relations(\alpha',t,t')$ or $\nu_\formulae(\phi,t')$.
As we proved, $\IB\models\nu_\relations(\alpha,t,t')$ implies that the first case is impossible.
This leaves the only alternative: $\nu_\formulae(\phi,t')\in\branch{B}$.

Therefore the general rule refinement condition
holds for $\Rfr(\ref{rule:box},T_{\K_m(\Not)}^+)$
and, by Corollary~\ref{corollary: refinement preserves constructive completeness}, 
$\Rfr(\ref{rule:box},T_{\K_m(\Not)}^+)$ is constructively complete.
\end{proof}

\begin{reftheorem}{theorem: tableau transformation: atoms}
Assume that
for an open branch $\branch{B}$ of the refined tableau $\Rf(\rho,T_L)$
and 
 for every set $Y$ of $\Lang{L}$-formulae from $\branch{B}$ the following holds.
\begin{trivlist}
 \item\textbf{\upshape Atomic rule refinement condition:}
    If all formulae in $Y$ are reflected in $\IB$
    then for every $E_1,\ldots,E_l\in Y$ and domain terms $t_1,\ldots,t_n$,
\begin{align*}
    \begin{aligned}
        & X_0(\seq{E},t_1,\ldots,t_n)\subseteq\branch{B}\ \text{implies that}\ \\
        & X_1(\seq{E},t_1,\ldots,t_n)=\{\Not\xi_1,\ldots,\Not\xi_k\}\text{and all $\xi_1,\ldots,\xi_k$ are \emph{\Lang{L}-atomic}}.
    \end{aligned}
\end{align*}
\end{trivlist}
Then, $\branch{B}$ is reflected in $\IB$.
\end{reftheorem}
\begin{proof}
We show the general rule refinement condition
holds in this case.
Assume 
$X_0(\seq{E},t_1,\ldots,t_n)$ is contained in $\branch{B}$ and 
        $\IB\not\models X_1(\seq{E},\ecl{t_1},\ldots,\ecl{t_n})$.
Therefore there is %
some $j=1,\ldots,k$
such that $\IB\models\xi_j(\ecl{t_1},\ldots,\ecl{t_n})$.
Since $\xi_j(t_1,\ldots,t_n)$ is \Lang{L}-atomic, by the definition of $\IB$
we have $\xi_j(t_1,\ldots,t_n)\in\branch{B}$.
Consequently, the rule $\rho_j$ has been applied in $\branch{B}$
to the set of premises $X_0(\seq{E},t_1,\ldots,t_n)\cup\{\xi_j(t_1,\ldots,t_n)\}$ 
and, hence, for some $i=2,\ldots,m$,
the set $X_i(\seq{E},t_1,\ldots,t_n)$ is contained in the branch $\branch{B}$.
Finally, by Theorem~\ref{theorem: tableau transformation1},
we have that $\branch{B}$ is reflected in $\IB$.
\end{proof}

\begin{reftheorem}{theorem:T_L^{c+}}
 $T_L^{c+}$ is sound and constructively complete for the logic $L$.
\end{reftheorem}
\begin{proof}
The rules \ref{rule:clause'} are 
particular cases of the \ref{rule:clause} rules for $k=m+n$.
That is, for any $m$ and $n$ such that $m+n>1$ there is a substitution which
converts the \ref{rule:clause} rule
with $k=m+n$ into the \ref{rule:clause'} rule. 

Furthermore, for any $k>1$ 
and a substitution $\sigma$
into the \ref{rule:clause} rule
there are $m$ and $n$ such that 
$m+n=k$ and,  under the substitution $\sigma$,
the premise and the conclusions of the rule \ref{rule:clause'}
coincide respectively with the premise and the conclusions of the \ref{rule:clause} rule
(modulo associativity and commutativity of the disjunction of $L$).

Therefore, the rules \ref{rule:clause} and the rules \ref{rule:clause'}
are derivable from each other. The theorem statement follows %
immediately.
\end{proof}

\end{document}